\documentclass[runningheads,a4paper]{llncs}

\usepackage{amsfonts,amssymb,amsmath}
\usepackage{natbib}

\usepackage{wrapfig}
\usepackage{llncsdoc}
\usepackage{epsfig,psfrag}
\usepackage{latexsym}
\usepackage{longtable}
\usepackage{tikz,pgf}
\usepackage{subfig}
\usepackage{wrapfig}

\usepackage{graphicx}
\usepackage{epstopdf}
\usepackage{color}

\newcommand{\bqn}{\begin{eqnarray}}
\newcommand{\eqn}{\end{eqnarray}}
\newcommand{\bq}{\begin{eqnarray*}}
\newcommand{\eq}{\end{eqnarray*}}
\usepackage[ruled,vlined]{algorithm2e}

\usepackage{url}
\usepackage{hyperref}
\hypersetup{colorlinks,%
citecolor=black,%
filecolor=blue,%
linkcolor=red,%
urlcolor=blue,%
pdftex}

\newcommand{\blue}[1]{{\color{blue} #1}}

\begin{document}

\title{Introduction to Logistic Regression}
 \titlerunning{Logistic Regression}
 
\author{Moo K. Chung
 }
\institute{
University of Wisconsin-Madison, USA\\
\vspace{0.3cm}
\blue{\tt mkchung@wisc.edu}
}
\authorrunning{Chung}

\maketitle

\begin{center}
July 29, 2020
\end{center}

\pagenumbering{arabic}

For random field theory based multiple comparison corrections 
In brain imaging, it is often necessary to compute the distribution of the supremum of a random field. Unfortunately, computing the distribution of the supremum of the random field is not easy and requires satisfying many distributional assumptions that may not be true in real data. Thus, there is a need to come up with a different framework that does not use the traditional statistical hypothesis testing paradigm that requires to compute $p$-values.  With this as a motivation, we can use a different approach called the {\em logistic regression}  that does not require computing the $p$-value and still be able to localize the regions of brain network differences  \citep{flury.1997, hastie.2003, chung.2008.MMBIA}. Unlike other discriminant and classification techniques that  tried to classify preselected feature vectors, the method here  does not require any preselected feature vectors and performs the classification at each edge level \citep{higdon.2004, shen.2004, thomaz.2006}.

\section{Logistic regression}
\index{logistic regression}
\index{connectivity!logistic regression}
\index{connectivity!probabilistic}

Logistic regression is useful for setting up a probabilistic model on the strength of brain connectivity and perform classification \citep{subasi.2005}. Suppose $k$ regressors are given for the $i$-th subject. These are both imaging and nonimaging phenotypes such as gender, age, education level and memory test score. Let $x_{i1}, \cdots, x_{ik}$ denote the measurements for the $i$-th subject. Let the response variable $Y_i$ be the probability of connection at a given edge, which is modeled as a Bernoulli random variable with parameter $\pi_i$, i.e., 
$$Y_i \sim \mbox{Bernoulli} (\pi_i).$$ 
$Y_i=0,1$ indicates the edge connected (assigned number 1) or disconnected (assigned number 0) respectively. 
 $\pi_i$  is then  the likelihood (probability) of the edge connected, i.e. $\pi_i = P(Y_i = 1)$.

Now consider linear model 
 \bqn Y_i = {\bf x}_i^{\top}\beta + \epsilon_i, \label{eq:logistic1} \eqn
where ${\bf x}_i^{\top}= (1,x_{i1},\cdots,x_{ik})$ and $\beta^{\top}=(\beta_0,\cdots ,\beta_k)$. We may assume 
$$\mathbb{E} \epsilon_i =0, \quad \mathbb{V} \epsilon_i = \sigma^2.$$
However, linear model (\ref{eq:logistic1}) is no longer appropriate since $$\mathbb{E}
Y_i = \pi_i = {\bf x}_i^{\top}\beta$$ but ${\bf x}_i^{\top} \beta$ may not be in the range $[0,1]$. The inconsistency is caused by trying to match continuous variables $x_{ij}$ to categorical variable $Y_i$ directly.  To address this problem, we
introduce the {\em logistic regression function}
$$t \to \frac{\exp t}{1 + \exp t} = \frac{1}{1 + \exp (-t)}  $$
  that links the response variable $\pi_i$ to the explanatory variables:
\bqn \pi_i &=&  \frac{\exp ({\bf x}_i^{\top}\beta_i)}{1+\exp ({\bf x}_i^{\top}\beta_i )} =   \frac{1}{1+\exp (-{\bf x}_i^{\top}\beta_i )}   \label{eq:logistic2}.
\eqn
Similarly we also have 
$$ 1 - \pi_i = \frac{1}{1+\exp ({\bf x}_i^{\top}\beta_i )}.$$
Using the {\em logit function}, we can rewrite (\ref{eq:logistic2}) as
$$\mbox{logit} (\pi_i) = \log \frac{\pi_i}{1-\pi_i} = {\bf x}_i^{\top}\beta_i.$$
Note
\bqn \frac{\partial \pi_i}{\partial \beta} =  {\bf x}_i \pi_i( 1 -\pi_i), \quad 
\frac{\partial }{\partial \beta}  \log(1 - \pi_i) = - {\bf x}_i \pi_i \label{eq:trick}\eqn
which can be used in simplifying the expression involving the gradient of loglikelihood.

\section{Maximum likelihood estimation}

The unknown parameters $\beta$ are traditionally estimated via the maximum likelihood estimation (MLE) over $n$ subjects at each connection in the brain network. The likelihood function based on the product of Bernoulli distributions is
\bq L(\beta|y_1,\cdots,y_n) &=& \prod_{i=1}^n \pi_i^{y_i}(1-\pi_i)^{1-y_i} \eq
The loglikelihood function is given by
\bqn \log L(\beta) &=& \sum_{i=1}^n y_i \log \pi_i + (1-y_i)\log(1-\pi_i) \label{eq:CE}\\
&=&  \sum_{i=1}^n y_i  \log \frac{\pi_i}{1-\pi_i}  + \log(1-\pi_i)\\
&=&  \sum_{i=1}^n y_i {\bf x}_i^{\top} \beta + \log(1-\pi_i).\eqn
Note ({\ref{eq:CE}) is sometime called the {\em cross entropy} \citep{bishop.2006}. 
From (\ref{eq:trick}), the maximum of loglikihood is obtained when its gradient $g$ vanishes:
\bqn g = \frac{\partial \log L(\beta)}{\partial \beta} = \sum_{i=1}^n {\bf x}_i (y_i - \pi_i)= X^{\top} ({\bf y} -  \boldsymbol{\pi}) =0, \eqn
where $\boldsymbol{\pi} = (\pi_1, \cdots, \pi_n)^{\top}$ and $X^{\top} = [{\bf x}_1, \cdots {\bf x}_n]$ is $(k+1) \times n$ data matrix of explanatory variables. Note the first row of $X^{\top}$ should be 1 corresponding to constant $\beta_0$ in the model.

Many computational packages such as R and MATLAB have the logistic regression model fitting procedure. Even SPM package widely used in brain imaging has the routine. Although we do not have the explicit formulas for the MLE, using the asymptotic normality of the MLE, the distributions of the estimators can be approximately determined. For large sample size $n$, the distribution of $\widehat \beta$ is approximately multivariate normal with means $\beta$ with the covariance matrix $H(\widehat \beta)^{-1}$, where $H$ is Hessian given by
\bqn H = \frac{\partial^2 \log L(\beta)}{\partial \beta ^{\top} \partial \beta} 
= \sum_{i=1}^n {\bf x}_i \frac{\partial}{\partial \beta} (y_i - \pi_i) 
&=& - \sum_{i=1}^n \pi_i(1-\pi_i){\bf x}_i {\bf x}_i^{\top} \nonumber \\
&=& -X^{\top} S X,
\eqn
where $S = diag(\pi_i(1 - \pi_i))$ is a diagonal matrix. Since the Hessian is a quadratic form with positive diagonal entires for $0 < \pi_i <1$, the loglikelihood function is concave and has one maximum. Finding such maximum is not hard  making the logistic regression very robust. 

\section{Newton's method}
The Newton-Raphson type of algorithms can be used to find the MLE in an iterative fashion. 
Consider the update of estimation: 
$$\beta^{j+1} = \beta^j + \Delta \beta.$$
The loglikelihood function can be expanded using the Taylor expansion around the $j$-th guess $\beta^j$:
$$\log L(\beta^j + \Delta \beta) = \log L(\beta^j)  + g^{\top} (\beta^j) \Delta \beta  + \frac{1}{2} \Delta \beta ^{\top} H(\beta^j) \Delta \beta.$$
The maximum of the loglikelihood is achieved when its derivative with respect to $\Delta \beta$ vanishes:
$$\frac{d \log L(\beta^j + \Delta \beta) }{d \Delta \beta} =0.$$
 Solving the equation, we get
$$ \Delta \beta =  - H(\beta^j)^{-1} g(\beta^j).$$
Thus, starting from initial guess $\beta^0$, we estimate $\beta$ iteratively as
\bqn \beta^{j+1} = \beta^j   - H(\beta^j)^{-1} g(\beta^j). \label{eq:betaupdate} \eqn
This 2nd order approximation is known to converge faster than the gradient descent method. 
In the matrix form, (\ref{eq:betaupdate}) can be written as
\bqn \beta^{j+1} = \beta^j + (X^{\top} S_k X)^{-1} X^{\top} ({\bf y} - \pi_k), \eqn
which is often known as {\em iteratively reweighed least squares} (IRLS) \citep{bishop.2006, murphy.2012}. 
Given group labels ${\bf y}=(y_1, \cdots, y_n)^{\top}$, the reasonable initial guess  is to start with $\beta^0=0$ and $\pi^0 = 1/2$ in all the entries. Given data matrix {\tt X} and group label vector {\tt y} the following MATLAB code estimates $\beta$ as {\tt beta} quickly.

\begin{verbatim}
[n k] = size(X);    
X=[ones(n,1) X];                          %constant 1 added
beta = zeros(k+1,1);                      %initial estimate
gnorm=1;                                  %size of gradient

while gnorm>0.001  
    extb = exp(X*beta);                   % equation (2) 
    pi = extb./(1 + extb);                           
    g = X'*(y - pi);                      % equation (7): gradient     
    gnorm = norm(g);        
    S=diag(pi .* (1 - pi));  
    H = X'*S*X;                           %equation (8): Hessian
    beta = beta + pinv(H)*g;              %equation (10)                                                               
end
\end{verbatim}

\section{Best model selection}

Consider following full model:
$$\mbox{logit} (\pi_i) = \beta_0 + \beta_1 x_1 + \beta_2 x_2 + \cdots + \beta_p x_p.$$ 
Let $\beta^{(1)} = (\beta_0, \cdots, \beta_q)^{\top}$ and $\beta^{(2)} = (\beta_{q+1}, \cdots, \beta_p)^{\top}$. The parameter $\beta^{(1)}$ corresponds to the parameters of the {\em reduced model}. Then we are interested in testing $$H_0: \beta^{(2)} = 0.$$ Define the {\em deviance} D of a model as
$D = -2 \log L(\widehat{\pi})$ which is distributed asymptotically as $\chi_{n-p-1}^2$.  Let $\widehat{\pi}^{(p)}$ and $\widehat{\pi}^{(q)}$ be the estimated success probabilities for the full and reduced models, and let $D_p$ and $D_q$ be the associated deviances. Then the log-likelihood ratio statistic for testing $\beta^{(2)}=0$ is 
$$2[\log L(\widehat{\pi}^{(p)}) - \log L(\widehat{\pi}^{(q)})]= D_q - D_p \sim \chi_{p-q}^2.$$ 

\begin{figure}[th!]
\begin{center}
\includegraphics[width=0.6\linewidth]{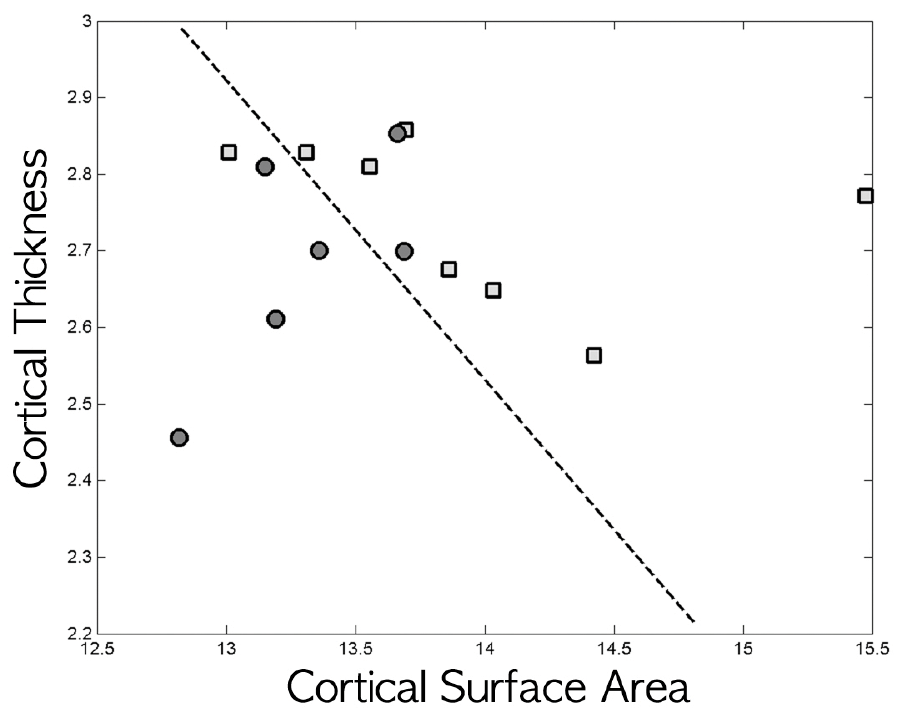}
\end{center}
 \caption{Based on 8 elderly controls (EC) (square) and 6 mild cognition impairment (MCI) (circle) subjects, a logistic discrimination analysis was performed using the average cortical thickness and total outer cortical surface area \citep{chung.2005.NI}. Using the cortical thickness alone results in 64.3$\%$ misclassification rate. On the other hand, using the both thickness and area results in significantly smaller  misclassification rate of 28.6$\%$. The rate is computed under the leave-one-out cross-validation scheme.  
 This shows that the univariate analysis based on cortical thickness alone is not sufficient to discriminate between the groups. On the other hand, analyzing data with cortical surface area reduces the error rate by 36$\%$. Instead of performing many different univariate analyses, doing a single multivariate analysis can be a more effect way of discriminating the two groups. The dotted line in the figure is the classification boundary. The upper part is EC while the lower part is MCI. The analysis shows that EC has larger cortical surface area and cortical thickness consistent with previous literature on AD. The data is used in this example came from Sterling C. Johnson of University of Wisconsin-Madison.} \label{fig:harmonics-logistic}
\end{figure}

\section{Logistic classifier}
\index{classification accuracy}
Discriminant analysis resulting from the estimated logistic model is called the {\em logistic discrimination}.
We classify the $i$-th subject according to a {\em classification rule}. The simplest rule is to assign the $i$-th subject as group 1 if
$$P(Y_i=1) > P(Y_i =0).$$
This statement is equivalent to $\pi_i > 1/2$. Depending on the bias and the error of the estimation, the value $1/2$ can be adjusted. For the fitted logistic model, we classify the $i$-th subject as group 1 if
${\bf x}_i^{\top}\beta_i > 0$ and as 0 if ${\bf x}_i^{\top}\beta_i < 0$. The plane ${\bf x}_i^{\top}\beta =0$ is the {\em classification boundary} that separates two groups. 
 Figure \ref{fig:harmonics-logistic} displays an example of classifying elderly controls (EC) from mild cognition impairment (MCI) subjects using the average cortical thickness and total cortical surface area.

\begin{figure}[t]
\centering
\includegraphics[width=0.7\linewidth]{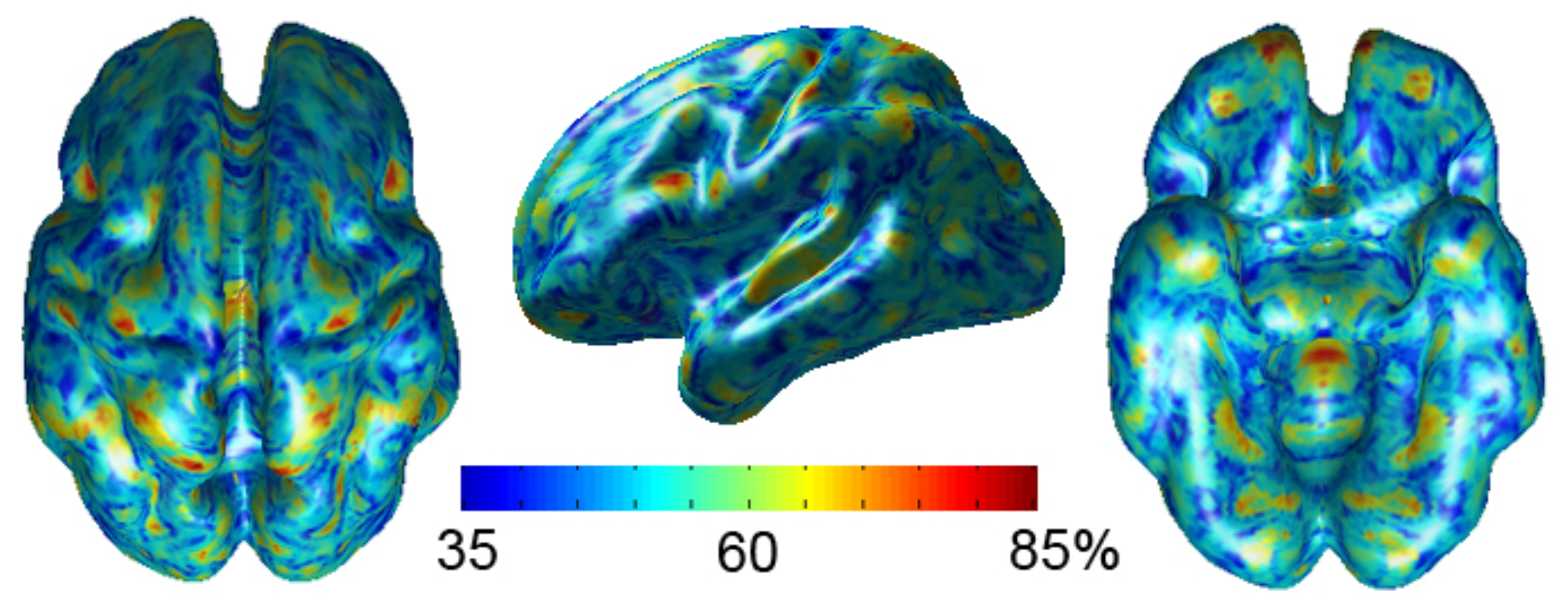}
 \caption{The discriminant power map on cortical thickness asymmetry pattern difference in 16 high functional autistic subjects and 12 normal controls \citep{chung.2008.MMBIA}. In this study, high functioning autistic subjects have a cortical thickness asymmetry pattern that differs reliably from controls. The discriminant power ranges from 32.1 to 85.7$\%$. The logistic discriminant analysis framework provides an alternative to the traditional corrected $p$-value approach in localizing signal differences the two group comparison setting.} \label{fig:power}
\end{figure}

The performance of classification technique is measured by the {\em error rate} $\gamma$, the overall probability of misclassification. The {\em cross-validation} is often used to estimate the error rate. This is done by randomly partitioning the data into the training and the testing sets. In the {\em leave-one-out} scheme, the training set consists of $n-1$ subjects while the testing set consists of one subject. Suppose the $i$-th subject is taken as the test set. Then using the training set, we determine the logistic model. Using the predicted model, we test if the $i$-th subject is correctly classified. The error rate obtained in this fashion is denoted as $e_{-i}$.
Note that $e_{-i} =0$ if the subject is classified correctly while $e_{-i}=1$ if the subject is misclassified.
The {\em leave-one-out error rate} is then given by 
$$\widehat \gamma = \frac{1}{n}\sum_{i=1}^n e_{-i}.$$
The {\em discriminant power} is then given as $1-\widehat \gamma$.
Figure \ref{fig:power} shows a study showing  the localization of abnormal asymmetry patterns in autistic subjects using the discriminant power computed at each surface mesh vertex.\\

\section{How classification accuracy is related to $p$-value} 
To formally test the statistical significance of the discriminant power, we use Press's Q statistic \citep{hair.1998}, which is given by
$$n(2\gamma   -1)^2 \sim \chi_1^2.$$ 

Press's Q statistic is asymptotically distributed as $\chi^2$ with one degree of freedom. Figure \ref{fig:Qplot} shows the plot of $p$-value of Press's Q-statistic as a function of discriminant power. Larger discriminant power should correspond to smaller $p$-value. For instance, for $n=28$ subjects, the discriminant power of 0.85 can correspond to the extremely small $p$-value of 0.0002. Matlab code below is used to plot Figure \ref{fig:Qplot}.
\begin{verbatim}
n=28
for i=1:1000
    power(i) = i/1000;
    pval(i)=1-chi2cdf(n*(2*power(i)-1)^2,1);
end;
plot(power,pval);
\end{verbatim}

\begin{figure}[t]
\centering
\includegraphics[width=0.7\linewidth]{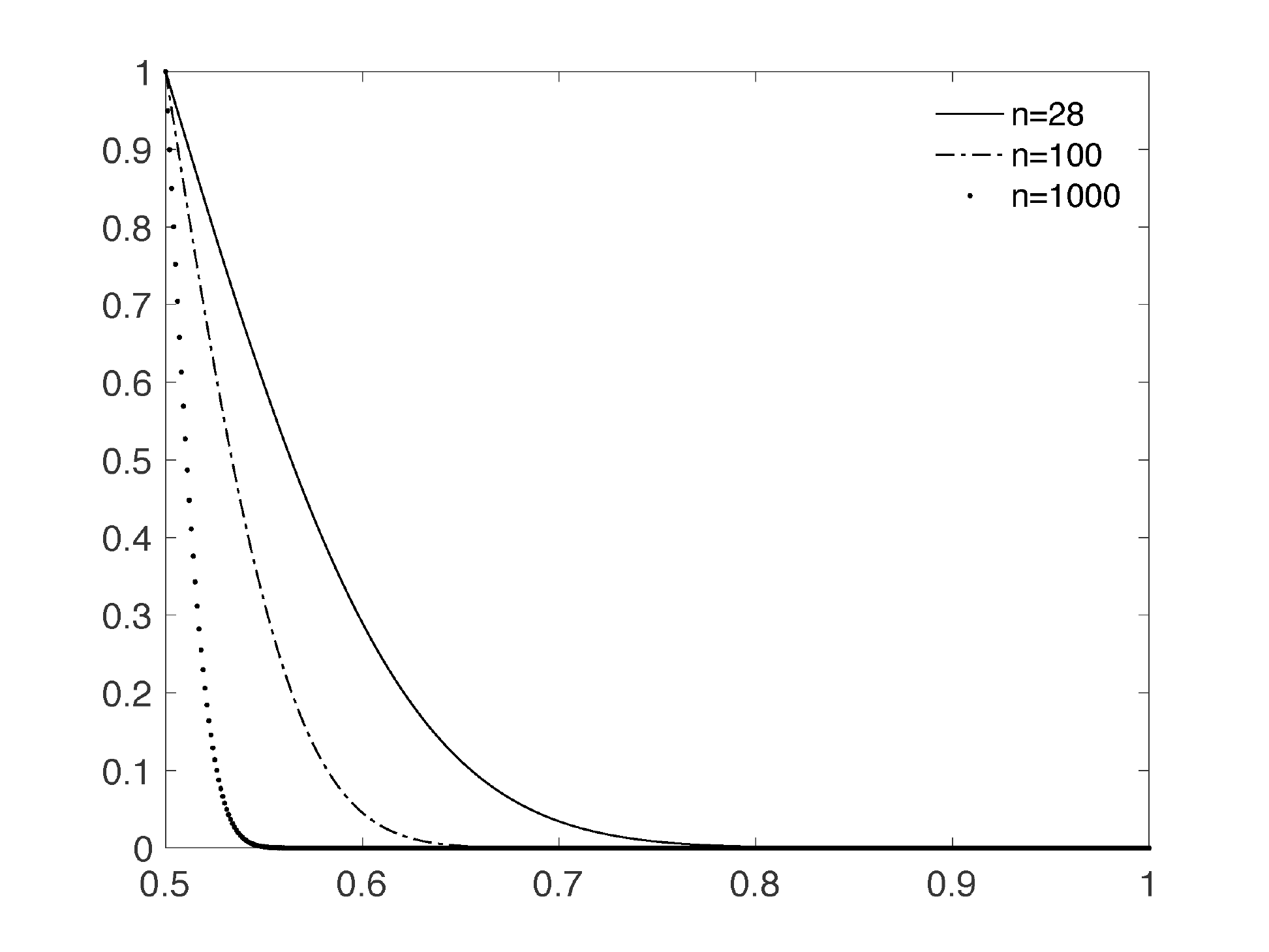}
 \caption{The $p$-value plot of Press's Q-statistic as a function of discriminant power for various sample sizes ($n$=28, 100, 1000) \citep{chung.2008.MMBIA}. For $n=28$, the discriminant power 0.85 corresponds to the small $p$-value of 0.0002.} 
 \label{fig:Qplot}
\end{figure}

To account for multiple comparisons, this small $p$-value needed to be corrected by computing the probability of the supremum distribution of a test statistic. However, this is not so trivial and requires the random field theory \citep{worsley.1995,worsley.1996}. This is left as a future study.

\section*{Acknowledgements}
The part of this study was supported by NIH grants NIH R01 EB022856 and R01 EB028753. We would like to thank Sterling C. Johnson of University of Wisconsin-Madison for providing the data used in Figure 1. We also like to thank Botao Wang of Xi'an Jiaotong University, China for pointing out the sign error in the original equation (8), which he corrected to be negative. The corresponding error in the MATLAB has been corrected. 
\bibliographystyle{agsm} 
\bibliography{reference.2020.07.22}

@book{bishop.2006,
  title={Pattern recognition and machine learning},
  author={Bishop, C.M.},
  year={2006},
  publisher={springer}
}

@article{chung.2005.NI,
  author =       {Chung, M.K. and Robbins, S. and Dalton, K.M. and Davidson, R.J. and Alexander, A.L. and Evans, A.C.},
  title =        {Cortical thickness analysis in autism with heat kernel smoothing},
  Journal =      {NeuroImage},
  volume =       {25},
  pages =        {1256--1265},
    year =         {2005}
}

@conference{chung.2008.MMBIA,
  title={{Quantifying cortical surface asymmetry via logistic discriminant analysis}},
  author={Chung, M.K. and Kelley, D.J. and Dalton, K.M. and Davidon, R.J.},
  booktitle={IEEE Computer Society Conference on Computer Vision and Pattern Recognition Workshops},
  pages={1--8},
  year={2008}
}

@book{flury.1997,
  author =       {Flury, B.},
  title =        {A First Course in Multivariate Statistics},
  PUBLISHER =    {Springer},
  year =         {1997},
}

@book{hair.1998,
  author =       {Hair, J.F. and Tatham, R.L and Anderson, R.E. and Black, W.C.},
  title =        {Multivariate Data Analysis},
  publisher =      {Prentice Hall, Inc.},
  year =         {1998},
}

@book{hastie.2003,
  author =       {Hastie, T., and Tibshirani, R., and Friedman, J.},
  title =        {The elements of statistical learning},
  publisher =      {Springer},
  year =         {2003}
}

@article{higdon.2004,
  author =       {R. Higdon, and N.L. Foster, and R.A. Koeppe, and C.S. DeCarli, and W.J. Jagust, and C.M. Clark, and N.R. Barbas, and S.E. Arnold, and R.S. Turner, J.L. Heidebrink, and S. Minoshima},
  title =        {A comparison of classification methods for differentiating fronto-temporal dementia from Alzheimer's disease using {FDG-PET} imaging},
  Journal =      {Stat Med},
  year =         {2004},
  volume =       {23},
  pages =        {315--326}
}

@book{murphy.2012,
title	= {Machine learning: a probabilistic perspective},
author	= {Murphy, K.P.},
year	= {2012},
publisher = {The MIT Press},
address	= {Cambridge, MA}
}

@article{shen.2004,
  author =       {Shen, L. and Ford, J. and Makedon, F. and Saykin, A.},
  title =        {Surface-based approach for classification of 3D neuroanatomical
structures},
  Journal =      {Intelligent Data Analysis},
  year =         {2004},
  volume =       {8},
  pages={519--542}
}

@article{subasi.2005,
  title={Classification of {EEG} signals using neural network and logistic regression},
  author={Subasi, A. and Ercelebi, E.},
  journal={Computer methods and programs in biomedicine},
  volume={78},
  pages={87--99},
  year={2005}
}

@conference{thomaz.2006,
  author =   {C. Thomaz, and J.P. Boardman, and S.J. Counsell, and D.L.G. Hill, and J.V. Hajnal, and A.D. Edwards, and M.A. Rutherford, and D. Gillies, and D. Rueckert},
  title =    {A Whole Brain Morphometric Analysis of Changes Associated with Preterm Birth},
  booktitle =    {SPIE Medical Imaging 2006: Image Processing},
  volume =       {6144},
  pages		= {1903--1910},
  year =     {2006},
}

@article{worsley.1995,
  author =       {K.J. Worlsey and J-B. Poline and A.C. Vandal and K.J. Friston},
  title =        {Test for Distributed, Non-focal Brain Activations},
  Journal =      {NeuroImage},
  year =         {1995},
  volume =       {2},
  pages =        {173-181}
}

@article{worsley.1996,
  author="K.J. Worsley and S. Marrett and P. Neelin and A.C. Vandal and K.J. Friston and A.C. Evans",
  title="A unified statistical approach for determining significant signals in images of cerebral activation",
  year="1996",
  Journal="Human Brain Mapping",
    VOLUME="4",
    PAGES="58-73"
}

\end{document}